\documentclass[a4paper,fleqn,usenatbib]{mnras}

\usepackage{graphicx}
\usepackage{subfig}
\usepackage{amsmath}
\usepackage{amssymb}
\usepackage[T1]{fontenc}
\usepackage{ae,aecompl}
\usepackage{graphicx}
\usepackage{newtxtext,newtxmath}

\title{The Chaotic Nature of TRAPPIST-1 Planetary Spin States}
\author[Vinson, Tamayo, \& Hansen]{
	Alec M. Vinson,$^{1}$\thanks{E-mail:vinson@astro.ucla.edu}, Daniel Tamayo$^2$,
	\& Brad M. S. Hansen$^{1}$
	\\	
	$^{1}$Mani L. Bhaumik Institute for Theoretical Physics, Department of Physics and Astronomy, University of California, Los Angeles, CA, 90095, USA\\
	$^{2}$Department of Astrophysical Sciences, Princeton University, Princeton, NJ, 08544, USA
}

\date{Accepted XXX. Received YYY; in original form ZZZ}

\pubyear{2019}

\begin{document}
	
	\label{firstpage}
	\pagerange{\pageref{firstpage}--\pageref{lastpage}}
	
	\maketitle
	
	\begin{abstract}		
	The TRAPPIST-1 system has 7 known terrestrial planets arranged compactly in a mean motion resonant chain around an ultra-cool central star, some within the estimated habitable zone. Given their short orbital periods of just a few days, it is often presumed that the planets are tidally locked such that the spin rate is equal to that of the orbital mean motion. However, the compact, and resonant, nature of the system implies that there can be significant variations in the mean motion of these planets due to their mutual interactions. We show that such fluctuations can then have significant effects on the spin states of these planets. In this paper, we analyze, using detailed numerical simulations, the mean motion histories of the three planets that are thought to lie within or close to the habitable zone of the system: planets d, e, and f. We demonstrate that, depending on the strength of the mutual interactions within the system, these planets can be pushed into spin states which are effectively non-synchronous. We find that it can produce significant libration of the spin state, if not complete circulation in the frame co-rotating with the orbit. We also show that these spin states are likely to be unable to sustain long-term stability, with many of our simulations suggesting that the spin evolves, under the influence of tidal synchronization forces, into quasi-stable attractor states, which last on timescales of thousands of years.
	\end{abstract}
	
	
	\section{Introduction}
	
	The search for habitable planets around nearby stars has spurred great interest in the lowest mass stellar hosts (see \cite{SBJ16} for a review of habitability prospects around M dwarfs), because terrestrial planets are physically larger and more massive relative to the host, and therefore easier to detect and characterize. Furthermore, the habitable zone is estimated to lie closer to the fainter low mass stars, making the chance of a transit detection greater \citep{C&D07}. For such reasons, the TRAPPIST-1 exoplanetary systems has sparked much interest in the scientific community, with seven planets observed in a compact configuration orbiting close to an ultra-cool dwarf star. Several of the reported planets may lie in a mean motion resonant chain, and several lie near or within the estimated habitable zone \citep{TRAP1}. 
	
	With orbital periods on the order of just a few days, tidal interactions between the planets and the host star are expected to be significantly stronger than those the Earth experiences from the Sun. This would seem to suggest that these planets, or any planets orbiting within the habitable zone around such small and cool stars, would almost certainly be tidally locked into a synchronous spin state \citep{KWR93, TBL18}, wherein the planet's spin period and orbital period are equal to each other. This is often considered as a significant barrier to the habitability of such planets, because of the extreme temperature differences such conditions engender across the face of the planet, with predicted atmospheric collapse caused by the freezing of volatiles on the night side for planets with atmospheric compositions below certain threshold densities  \citep{KWR93, JHR97, WFS11, Wordsworth15}.
	
	In previous work \citep{VH17},  we proposed a potential solution to this tidal locking problem. For a synchronously rotating body (like the Earth's moon), its spin rate equals its orbital rate so as to always present the same face toward the primary. However, planets in, or near, a mean motion resonance undergo mean motion variations. These variations create a moving target for the synchronicity of the spin rate to the mean motion, which can affect the spin in such a way that it stably librates about an equilibrium point in the frame co-rotating with the orbit, or else circulates completely, resulting in full, stable, stellar days. As case studies, we considered the TRAPPIST-1 system, as well as the Kepler candidate K00255 system, which contains a star of mass $\left( 0.53 \pm 0.06 \right) M_\odot$, a confirmed planet of radius $R = \left(2.51 \pm 0.3\right) R_\oplus$ and orbital period of 27.52 days, and a candidate companion planet that would be in a 2:1 mean motion resonance with a radius of $\left(0.68 \pm 0.08\right) R_\oplus$ and orbital period of 13.60 days. We found that timescales of libration or circulation were such ($\lesssim 1$ years for a TRAPPIST-1 inspired system, and $\lesssim 10$ years for a K00255 inspired system) that they were less than the atmospheric response time of $\lesssim 10$ years for Earth-like planets \citep{SMS}. Thus, this model provides a potential mechanism to prevent the predicted atmospheric collapse on otherwise synchronously rotating planets.
	
	In our previous paper, we addressed only how the spin of a planet would be affected by a single companion of much greater mass near resonance. In reality, planets in multi-planet systems tend to have similar masses \citep{Millholland17,Weiss18}, and thus mutually interact. Additionally, many planetary systems are highly multiple and can even form multi-resonant chains, with TRAPPIST-1 as the prime example. In this paper, we therefore incorporate the effects of multiple planets in a resonant chain on a planet's spin state, using the TRAPPIST-1 system to showcase our model and possible behaviors.
		
	In \S \ref{model} we reintroduce our spin model and describe how we incorporate within it orbital histories of TRAPPIST-1-like systems using N-body integrations from \cite{TRP17}. In \S \ref{results} we present our findings, and in \S \ref{discussion} discuss the significance and some implications of our results. \S \ref{conclusion} summarizes our findings and provides some concluding remarks.

	\section{Model and Methods}\label{model}
	
	We describe the planetary spins using the same basic framework presented in \cite{VH17}. Thus we will consider a system with a planet of mass $m$ orbiting a star of mass $M_*$, with zero obliquity between the planet's spin and its orbit. We let $A$, $B$, and $C$ be the principal moments of inertia of the planet with $C$ being the moment about the spin axis and $A$ being the moment about the long axis of the planet in the plane of the orbit such that $B > A$. We define $\theta$ to be the angle formed between the long axis of the planet and a stationary line in the inertial frame, and another angle $\gamma \equiv \theta - M$, where $M$ is the mean anomaly. $\gamma$ then roughly corresponds to the longitude of the substellar point for a planet on a near-circular orbit. We then apply the equation of motion for $\gamma$ as derived and presented in \cite{GP66,GP68} and \cite{MD}, under the assumption that $\dot{\theta} \simeq n$
	
	\begin{equation}
	\ddot{\gamma} + \textnormal{sgn}\left[H(e)\right] \frac{1}{2} \omega_S^2 \sin 2\gamma + \dot{n} = 0
	\label{eqn:spin_no_damp}
	\end{equation}
	
	\noindent where $n = \dot{M}$ is the mean motion, $H(e)$ is a power series in orbital eccentricity $e$, and $\omega_S^2 = 3 n^2 \left(\frac{B-A}{C}\right)|H(e)|$. 
	
	However, a crucial difference between equation \ref{eqn:spin_no_damp} presented here and that presented in other sources is that we do not dispose of the $\dot{n}$ term on the presumption that it should be insignificant, in which case one is left with the differential equation for a pendulum  with two stable equilibria in the variable $\gamma$. In fact, as shown in our previous work, the $\dot{n}$ term can provide a substantial driving that alters the evolution of $\gamma$ if there is a companion near a mean motion resonance which can induce strong mean motion oscillations.
	
	Previously \citep{VH17}, we applied the ``pendulum model'' presented in \cite{MD} to describe the $\dot{n}$ term in equation \ref{eqn:spin_no_damp}, wherein $n$ evolved as a simple pendulum under the effects of a single companion planet in resonance. Thus, equation \ref{eqn:spin_no_damp} described something analogous to a forced pendulum for the evolution of $\gamma$, with the simple case having two stable equilibria corresponding to opposite faces of the planet along the planet's long axis pointing to the host star. The main difference from a typical forced pendulum is that our forcing term itself behaved like a pendulum instead of the more typical case of a simple sinusoid. The forced pendulum is a popular topic in dynamical studies due to its interesting chaotic behaviors when the natural and driving frequencies are close in value. Indeed, after incorporating tidal damping effects in the model, our results in \cite{VH17} depicted a wide range of interesting limit cycles for example systems based on the TRAPPIST-1 and K00255 planets.
	
	However, many systems are found to be highly multiple, including the TRAPPIST-1 system that inspired much of our work. To have a fuller understanding of the possible behaviors of the spin, we must then expand the model to incorporate the effects of multiple resonant companions. This is the main difference between the work presented in this paper and our previous one: we now consider the effects of \textit{multiple}, mutually interacting companions in or near mean motion resonances in order to understand how our model applies to resonant-chain systems such as TRAPPIST-1, wherein there are seven planets compactly arranged into a long resonant chain. This introduces a more difficult problem in describing how exactly the mean motion of any particular planet varies in time, with interactions from many companions. We therefore use N-body integrations of the TRAPPIST-1 system by \cite{TRP17} to calculate the variations in the orbital eccentricities and mean motions numerically.
	
	TRAPPIST-1 presents a very attractive system as a compact analog of the inner Solar System, with the seven planets contained within the system receiving between about 10\% and 400\% the stellar irradiation of the Earth. As first announced by \cite{TRAP1}, the inner six planets in the system form the longest known near-resonant chain of exoplanets, with orbital period ratios of approximately 8/5, 5/3, 3/2, 3/2, and 4/3 between planets c, d, e, f, and g, respectively, and their nearest inner companion. We also note that the period ratios of of e to d, f to e, and g to f, suggest that the resonances are of first-order, which also suggest larger variations in $\dot{n}$ as can be seen in the model described in \cite{VH17}.

	\subsection{Tidal Damping}
	
	Tidal effects on stellar and planetary orbits is a long-studied subject (see \cite{Og14} for a review). Tidal forces between planets and their host star are expected to be exceptionally strong in systems such as TRAPPIST-1 due to the close proximity between the planets and the host star. In the simple unforced case, this damping would push planets into synchronous spin-orbit states. 
	
	There is growing literature on the mechanisms of tidal dissipation in Earth-like planets \citep{EL07,ME13,FM13,CBLR}, but we will use the simpler constant time lag formalism, which is sufficient for illustrative purposes and has the attractive quality that it avoids unphysical discontinuities by having tidal torque going to zero as $\dot{\gamma}$ goes to zero. In particular, we wish to focus on the spin behavior driven by changes in the forcing without contamination by spin flips associated with abrupt changes in the sign of the spin damping forces. The formalism we use to describe tidal dissipation in this work is the same as we used previously in \cite{VH17}.
	
	Based on the work of \cite{H81} and \cite{EKH98} on tidal dissipation, we use equations (4) and (14) of \cite{H10} to describe tidal dissipation on $\gamma$ with
	
	\begin{equation}
	\ddot{\gamma} = -\frac{15}{2} \dot{\gamma} \frac{M_*}{m} \left(\frac{R}{a}\right)^6 M_* R^2 \sigma
	\label{eqn:tidal_diss}
	\end{equation}
	
	\noindent where $R$ is the planetary radius and $a$ is the semi-major axis of the orbit. The strength of the dissipation is described in terms of a bulk dissipation constant $\sigma$. Since much of the literature phrases the discussion of tidal evolution in terms of the tidal dissipation factor $Q$, we choose to also use \cite{H10} to translate the bulk dissipation $\sigma$ into the equivalent tidal $Q$, while still formally using the aforementioned convention which describes tidal dissipation in terms of the bulk dissipation constant $\sigma$. We can then choose an equivalent tidal factor $Q$ under the specific forcing applied to Earth to describe the tidal damping strength in these Earth-like planets, converting to bulk dissipation constant with
		
	\begin{equation}
	\sigma = \dfrac{1}{2Q}\left(\dfrac{G}{\Omega R^5}\right)
	\end{equation} 
	
	\noindent wherein we set the tidal forcing frequency $\Omega = 2\pi / \textnormal{day}$, approximately equal to the tidal frequency on Earth by the Moon. 
	
	Adding equation \ref{eqn:tidal_diss} to equation \ref{eqn:spin_no_damp}, we get our full equation of motion
	
	\begin{equation}
	\ddot{\gamma} + \frac{1}{2}\omega_S^2 \sin 2\gamma + \dot{n} + \epsilon \dot{\gamma} = 0
	\label{eqn:full}
	\end{equation}
	
	\noindent where we let $\epsilon = \frac{15}{2}\frac{M_*}{m}\left(\frac{R}{a}\right)^6 M_* R^2 \sigma$ define the strength of the dissipation. We note that while the spin evolution depends on the orbital behavior through $\dot{n}$ and the eccentricity dependence in $\omega_S$, the spin has such negligible angular momentum compared to the orbit that the spin does not feed back on the orbital evolution. This allows us to simply plug in the orbital histories from N-body integrations to calculate the spin evolution.
	
	\subsection{Setup}\label{setup}
	
	To extract orbital parameters, we use the numerical integrations of \cite{TRP17}, which simulated a range of initial conditions near the observed resonant configurations of TRAPPIST-1. These spanned configurations at or near the centers of the observed resonant chain where eccentricities and mean motions remain nearly constant (i.e., low spin forcing), to ones near their separatrices where the orbits undergo larger, chaotic oscillations (i.e., strong spin forcing). These integrations thus can span a range of dynamical behaviors, but we note that other works have determined better fits to the actual TRAPPIST-1 system \citep[e.g.][]{TRAP1, GSG18}. However, we wish to illustrate behaviors for a wide range of possibilities that can be applied not only to TRAPPIST-1, but to similar systems as well.

	The integrations in \cite{TRP17} were performed using the {\tt WHFAST} integrator \citep{RT15} in the REBOUND N-body package \citep{RL12}, with a timestep of 7\% of the innermost planet's orbital period. Details of their initialization by migrating them into the observed resonant chain with parametrized disk forces can be found in \cite{TRP17}. We extracted their publicly available\footnote{\url{https://github.com/dtamayo/trappist}} {\tt SimulationArchives}, which allow for fast, parallel extraction of system parameters at arbitrary times \citep{RT17}, and sampled eccentricities and mean motions at a cadence of 10\% of the orbital period for each planet. With the mean motion and eccentricity histories, we then used spline interpolation so that we could extract values of mean motion $n$, $\dot{n}$, and eccentricity $e$, at arbitrary times to feed into equation \ref{eqn:full} and perform an adaptive time-step, Fourth-Order Runge-Kutta integration to find the evolution of the spin parameter $\gamma$.

	Choices also need to be made in regards to assumptions about the planets and their tidal damping strengths, which can result in different behaviors. We choose to calibrate equation \ref{eqn:tidal_diss} to two different damping strengths. Studies suggest a tidal parameter $Q \sim 10$ for the Earth, exerted primarily by pelagic turbulence in the oceans \citep{ER00}, so we will take this as our stronger damping estimate. For a water-poor terrestrial planet, we adopt $Q=100$ as our weaker damping, based on estimates for Mars \citep{VDP07}. We also choose a triaxiality factor of $\left(B-A\right)/C = 10^{-5}$ based on considerations of empirical evidence determined for the Earth, Mars, and Mercury  \citep{Chen15,EKE,Margot}. We note that reasonable assumptions on this triaxiality factor can vary among planets by up to an order of magnitude, and can thus have an effect on the natural libration frequency $\omega_S$ by up to a factor of $\sim 3$. We find, however, that our overall results do not change significantly within this range of reasonable triaxiality assumptions.

	Finally, we focus on the spins of planets d, e, and f (while under the influence of all other companions in the system). Much of our interest in this problem is motivated by habitability questions, and these three planets all lie within the estimated habitable zone \citep{KRK13,KWH16}. These planets, while likely to have experienced water loss during the host star's $\sim 1$ Gyr long pre-main sequence phase, are also likely to still have been able to retain significant oceans depending on initial water levels \citep{BSO17}.

	\section{Results} \label{results}
	
	We use various outcomes from different initial conditions of REBOUND simulations of the orbital parameters of the TRAPPIST-1 system, picked to span a range in forcing strength from negligible to strong (see Appendix \ref{appendix}), and use these as inputs for our own spin model detailed in this paper.  Many of the simulations are dominated by a primary frequency of the mean motion variations, with many others manifesting strong secondary frequencies. We use eighteen different simulations as input for our spin model, ranging from very low amplitude librations of the mean motion to higher amplitude variations, all of which are long-lived configurations \citep{TRP17}.
	
	In this paper we look at those planets closest to the estimated habitable zone: planets d, e, and f. With 18 different REBOUND simulations to apply to each of these planets, coupled with two different tidal strengths we choose to test ($Q=10$ and $Q=100$), we have a total of 36 different simulations of the spins of each of these three planets. We observe a variety of different behaviors that emerge from different integrations. These include integrations wherein the spin of the planet remains effectively purely synchronous, with very small librating amplitudes. Other simulations depict higher amplitude librations in the spin, while others even exhibit periods of complete circulation. However, most of the realizations we studied exhibit chaotic evolution, with the spin often alternating between periods of libration and circulation. There is a small subset of integrations with approximately regular orbital behavior that do stably librate over long timescales.

	\begin{figure}
		\centering
		\includegraphics[width=1.0\linewidth]{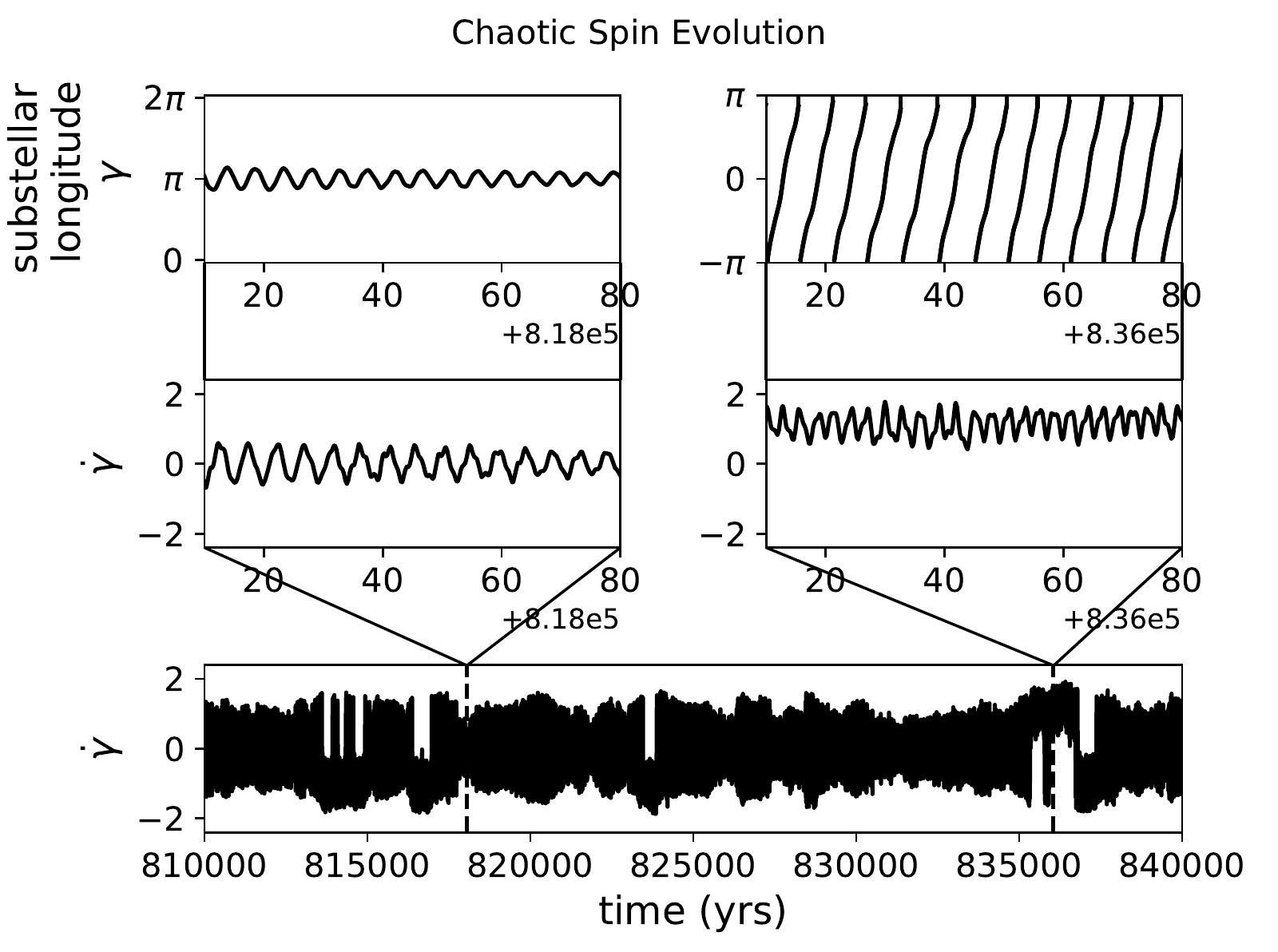}
		\caption{Example simulation of spin of planet f with $\dot{\gamma}$ plotted against time, wherein we set $Q=100$. Years are used for units involving time. The bottom panel depicts the  long-term evolution of $\dot{\gamma}$ over 30,000 years. The top four panels zoom in to depict two different types of behavior that occurs in this particular simulation, with the middle panels depicting evolution of $\dot{\gamma}$ and the top panels depicting evolution of $\gamma$. On the left panels we observe libration of $\gamma$ about $\pi$. On the right panels we observe a time when there is full circulation of $\gamma$, which can remain quasi-stable for approximately $10^3$ years. Overall, this simulation depicts a chaotic evolution of the spin state, switching among librating and circulating states.}
		\label{fig:runf18_gamma_dot_zoom}
	\end{figure}
	
	\begin{figure}
		\centering
		\includegraphics[width=1.0\linewidth]{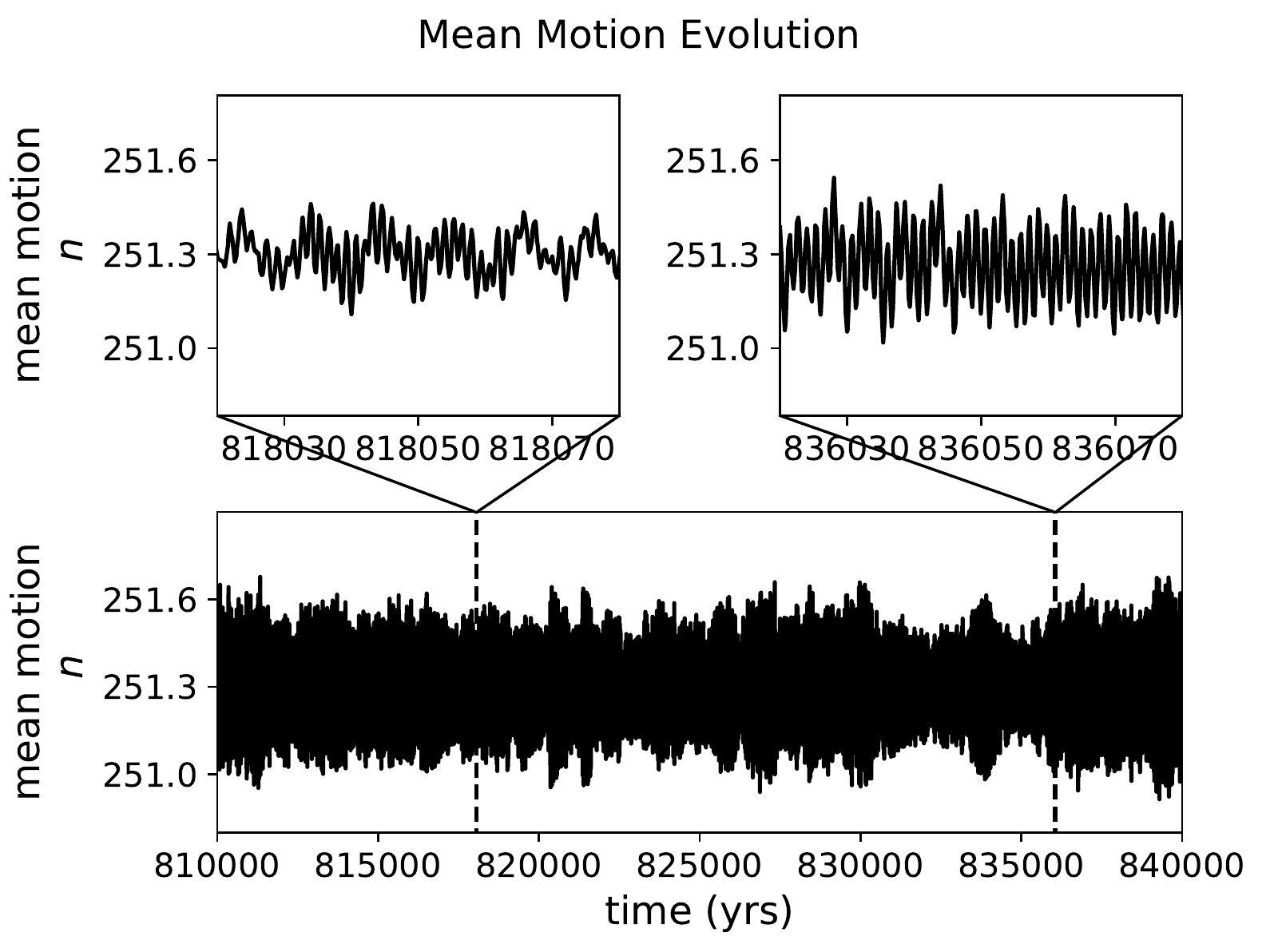}
		\caption{Example simulation of planet f depicting evolution of mean motion $n$ versus time taken from a REBOUND simulation. This is used as input for our spin model, a corresponding simulation of which is shown in Figure \ref{fig:runf18_gamma_dot_zoom}. The bottom panel presents long-term evolution of $n$, while the top two panels zoom in to depict behavior at different instances in time. We can compare this to Figure \ref{fig:runf18_gamma_dot_zoom} to see how the mean motion input affects the spin state. We find that the most dominant frequency when performing a Fourier Transform on $n$ is $\omega_M = 5.1$ yr$^{-1}$. We also find that the standard deviation of $\dot{n}$, which we can use as a proxy for overall forcing strength, is $\sigma(\dot{n}) = 0.55$ yr$^{-2}$. We note, however, that $\sigma(\dot{n})$ varies throughout the simulation, and we can observe here that the mean motion has larger variations in the top right panel than in the top left panel, corresponding to circulating and librating states for the spin argument $\gamma$ as seen in Figure \ref{fig:runf18_gamma_dot_zoom}, which is suggestive that the behavior of the spin depends strongly on the strength of mean motion variations. }
		\label{fig:runf18_n_zoom}
	\end{figure}

	\subsection{High Amplitude, Irregular Forcing}
	
	One representative example is presented in Figure \ref{fig:runf18_gamma_dot_zoom}, with variations in mean motion attained from the corresponding REBOUND simulation presented in Figure \ref{fig:runf18_n_zoom}. We can see a variety of behaviors in just this one simulation. First we notice that, due to the inherent chaos of the behavior of the orbital parameters from the input taken from the REBOUND simulation, the spin state also never reaches a true stable equilibrium. In this particular simulation, we see switching among different quasi-stable states, which each last on the order of a few thousand to tens of thousands of years. Some of these states include moderate-amplitude librations of $\gamma$ about either $0$ or $\pi$, while others are states of complete circulation, wherein we observe full stellar coverage on the surface of the planet.
	
	\subsection{Regular Forcing}
	
\begin{figure}
		\centering
		\includegraphics[width=1.0\linewidth]{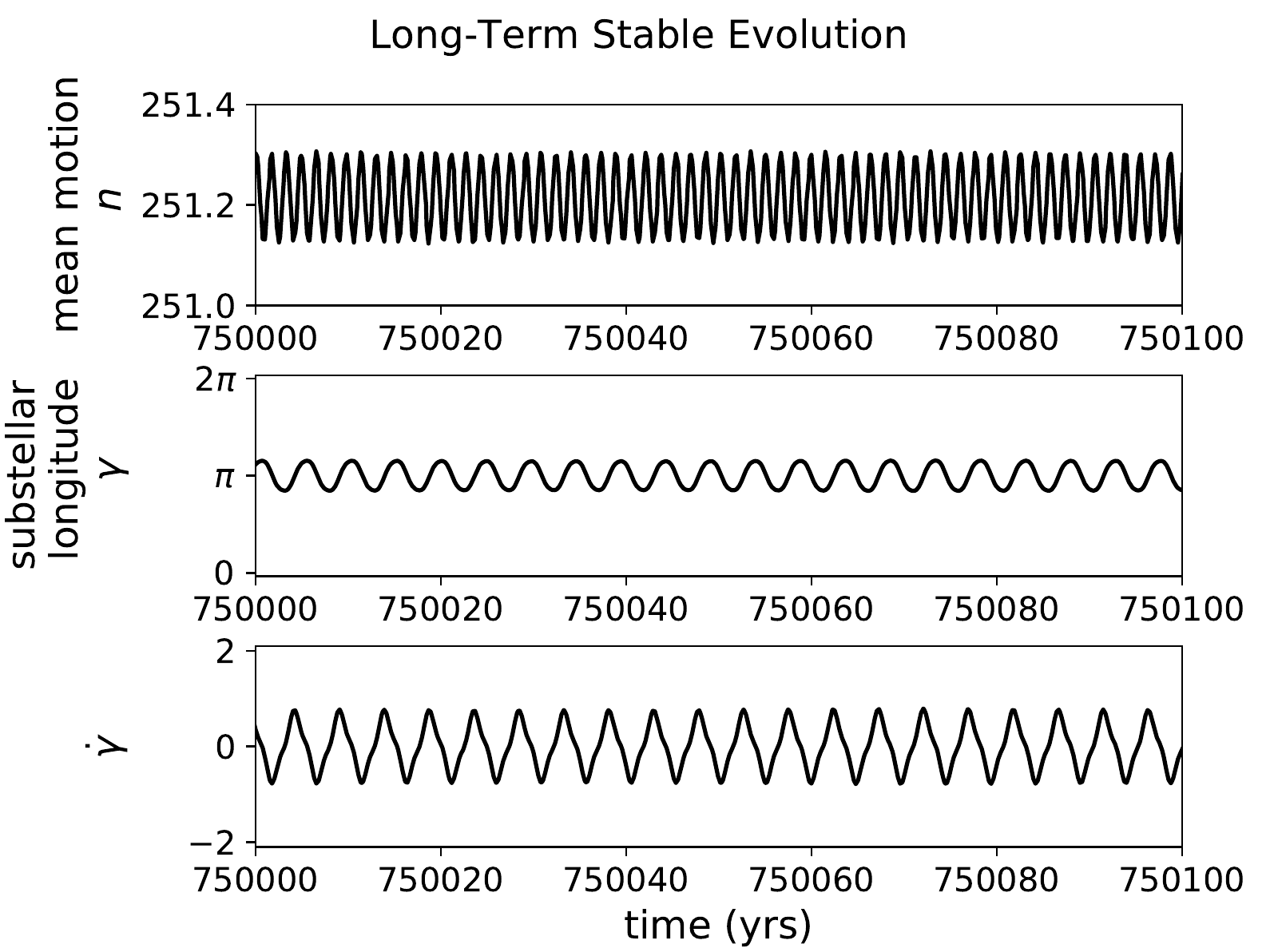}
		\caption{Example simulation of planet f depicting evolution of mean motion $n$ in the top panel, $\gamma$ in the middle panel, and $\dot{\gamma}$ in the bottom panel. This is an example of a simulation which has a very stable evolution in mean motion which is strongly dominated by just one frequency, $\omega_M = 4.18$ yr$^{-1}$, as shown in the Fourier transform in Figure \ref{fig:runf13_freq_spectrum}. Thus, we also observe a stable evolution of the spin argument $\gamma$, wherein we observe a stable libration of $\gamma$ about $\pi$. We also note that the standard deviation of $\dot{n}$, which we use as a proxy for forcing strength, is $\sigma(\dot{n}) = 0.26$ yr$^{-2}$. }
		\label{fig:runf13}
	\end{figure}

\begin{figure}
		\centering
		\includegraphics[width=1.0\linewidth]{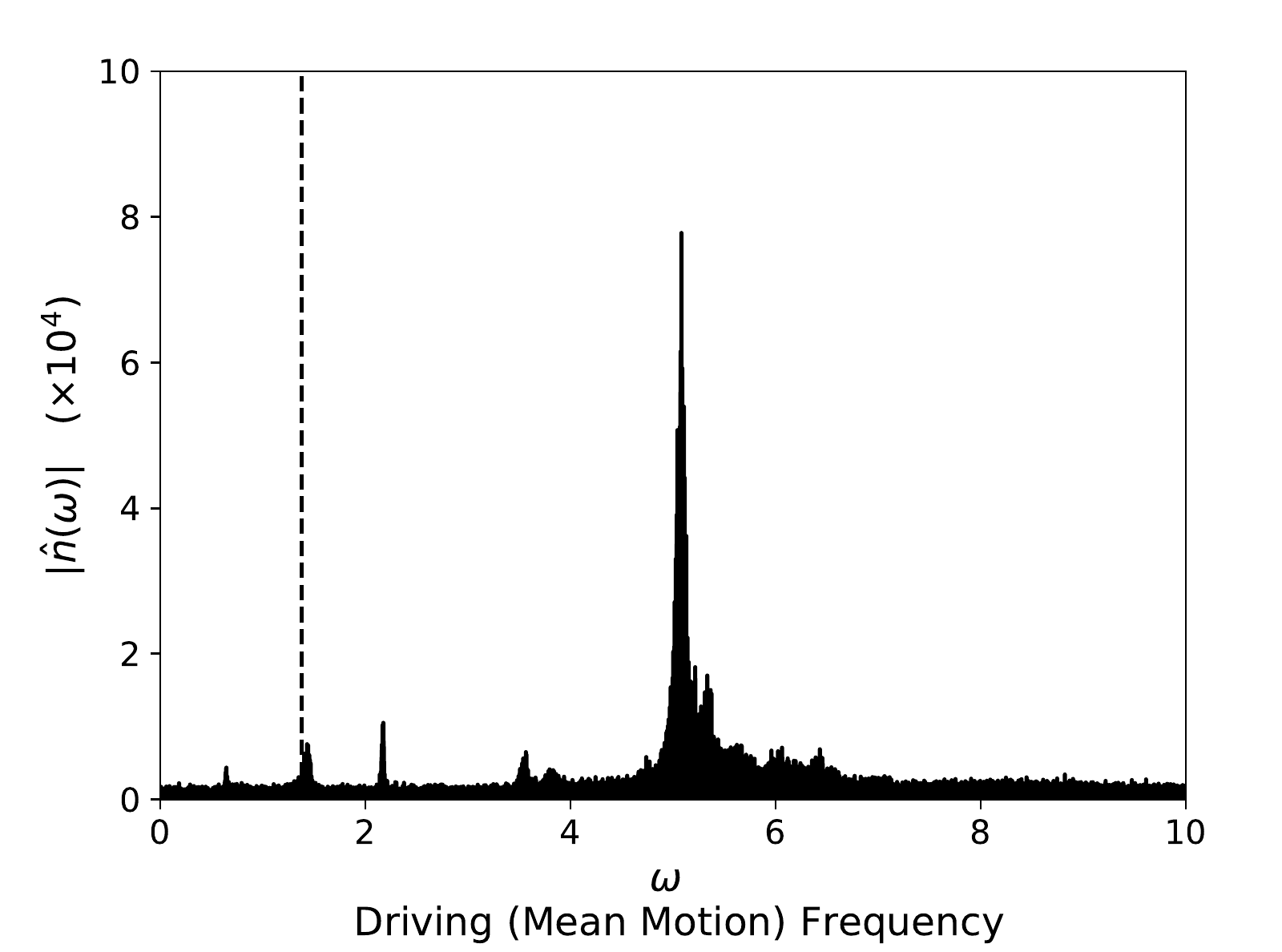}
		\caption{Absolute value of the frequency spectrum of the mean motion $n(t)$, denoted $|\hat{n}|$, for the entire time range of the simulation corresponding to Figure \ref{fig:runf18_n_zoom}. Here we see that, while there is a primary frequency which dominates in the behavior of $n(t)$, $\omega_M = 5.1$ yr$^{-1}$, there are also secondary frequencies which can change the behavior of $n$, and thus its influences on the spin argument $\gamma$. We also plot vertical dashed lines to indicate the spin frequency for natural libration for planet f, $\omega_S = 1.38$ yr$^{-1}$ for our assumed triaxiality  $\left(B-A\right)/C$. Here we note that $\omega_S$ happens to line up approximately with one of the secondary driving frequencies.}
		\label{fig:runf18_freq_spectrum}
	\end{figure}

\begin{figure}
		\centering
		\includegraphics[width=1.0\linewidth]{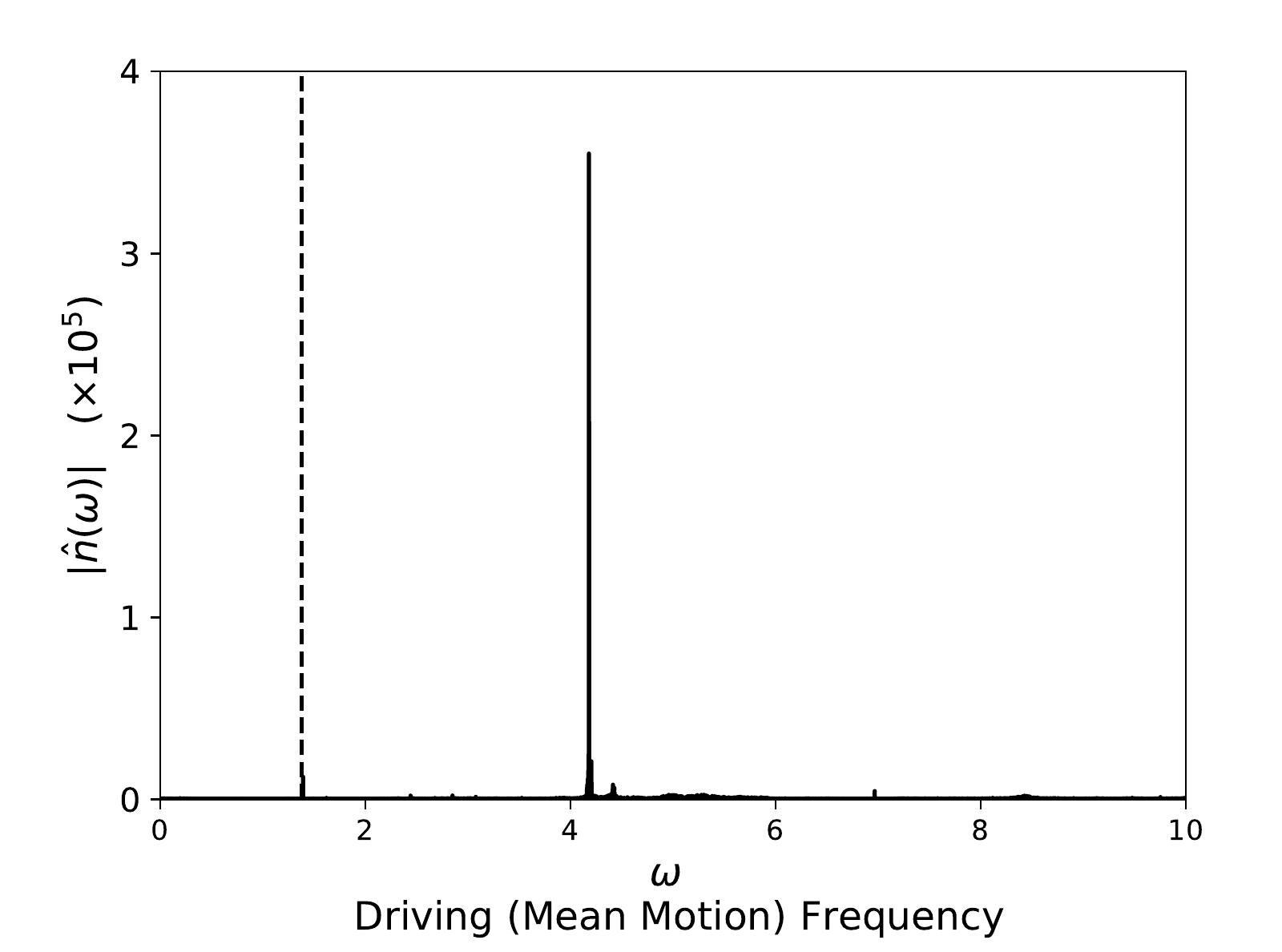}
		\caption{Absolute value of the frequency spectrum of the mean motion $n(t)$, denoted $|\hat{n}|$, for the entire time range of the simulation corresponding to Figure \ref{fig:runf13}. We also plot vertical dashed lines to indicate the natural libration frequency for planet f, $\omega_S = 1.38$ yr$^{-1}$ for our assumed triaxiality $\left(B-A\right)/C$. Here we see that the evolution of $n$ can largely be described in terms of just one dominant frequency, $\omega_M = 4.18$ yr$^{-1}$, which is approximately three times the natural libration frequency $\omega_S$. }
		\label{fig:runf13_freq_spectrum}
	\end{figure}

	Other simulations can depict far more regular behavior if the behavior of the driving (i.e. the mean motion) is also very stable. One such example is shown in Figure \ref{fig:runf13}, where we show the evolution of $\gamma$, $\dot{\gamma}$, and $n$, for a small period of time during the simulation (though the entire simulation depicts the same, stable behavior). We note the very regular behavior of the mean motion $n$, especially as compared to the evolution of $n$ for our other example shown Figure \ref{fig:runf18_n_zoom}. In fact, the example depicted in Figure \ref{fig:runf13} is among the closest to our classical case presented in our previous paper \citep{VH17}, where behavior of mean motion $n$ can largely be described as a simple pendulum with a single frequency. This can also be compared in Figures \ref{fig:runf18_freq_spectrum} and \ref{fig:runf13_freq_spectrum}, which depict the Fourier Transforms of the mean motion shown in Figure \ref{fig:runf18_n_zoom} and Figure \ref{fig:runf13}, respectively. Where Figure \ref{fig:runf18_freq_spectrum} exhibits multiple frequencies in the mean motion, in Figure \ref{fig:runf13_freq_spectrum} we see that the mean motion is dominated by just one. As a result of a more stable mean motion evolution, we observe a stable libration of $\gamma$ in the \ref{fig:runf13} with an amplitude of about $0.4$ radians about $\pi$.

	\begin{figure*}
		\centering
		\includegraphics[width=1.0\linewidth]{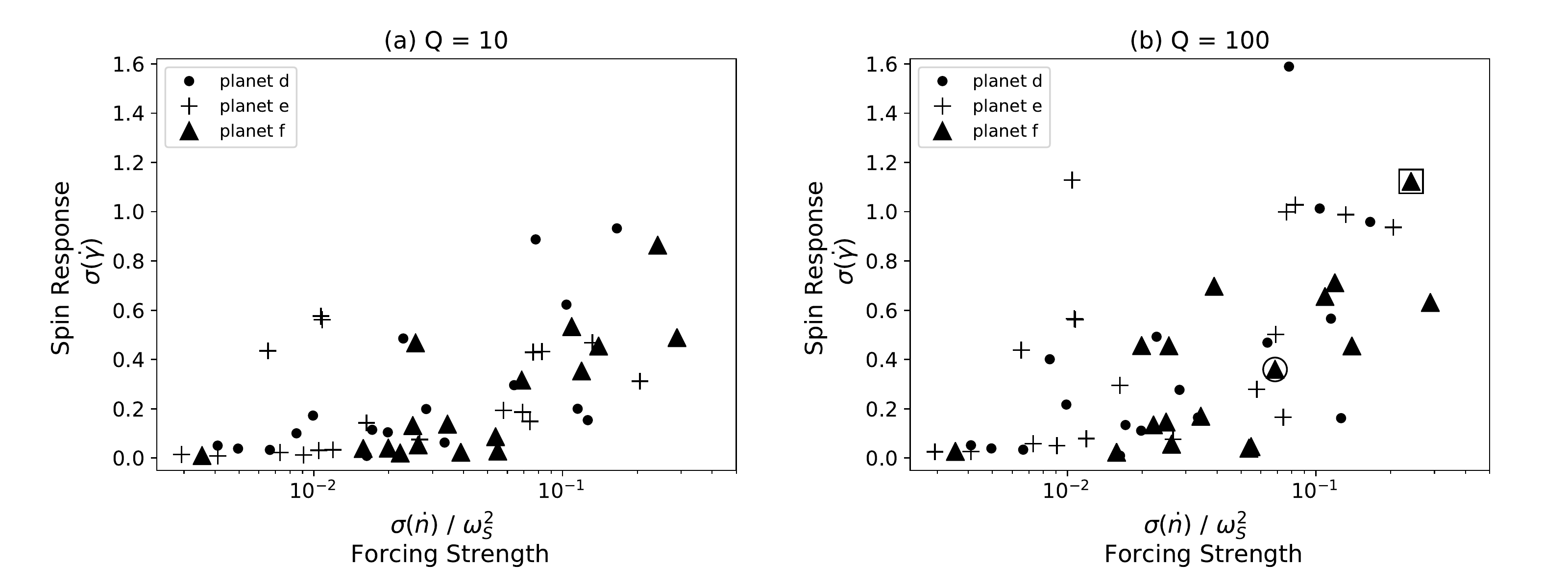}
		\caption{Standard deviation of $\dot{\gamma}$ versus standard deviation of $\dot{n}$ normalized to $\omega_S^2$ for each of our integrations with tidal factor $Q = 10$ (i.e. higher tidal damping) depicted in panel (a) on the left, and tidal factor $Q = 100$ depicted in panel (b) on the right. Integrations for planets d, e, and f, are represented by dots, plus signs, and triangles, respectively. The cases studied from Figures \ref{fig:runf18_gamma_dot_zoom} and \ref{fig:runf13} are enclosed in a square and a circle, respectively. We see an overall trend that variations in $\dot{\gamma}$ increase with variations in $\dot{n}$. Higher variations in $\dot{\gamma}$ indicate larger responses to the driving of the spin argument $\gamma$, and thus would suggest the potential for larger total stellar coverage. Years are used for units with time.}
		\label{fig:spin_response_vs_forcing}
	\end{figure*}

	\begin{figure*}
		\centering
		\includegraphics[width=1.0\linewidth]{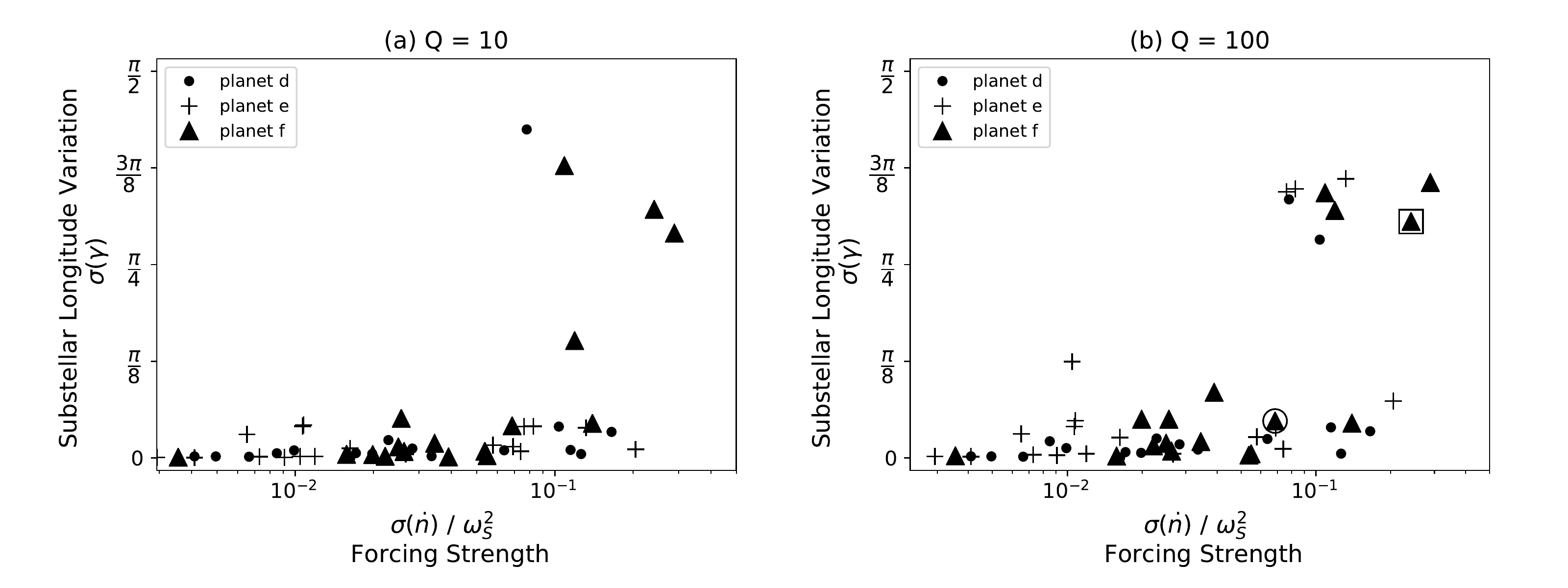}
		\caption{Standard deviation of $\gamma$ versus standard deviation of $\dot{n}$ normalized to $\omega_S^2$ for each integration with tidal parameter $Q = 10$ (i.e. higher tidal damping) being depicted in panel (a) on the left and tidal factor $Q = 100$ being depicted in panel (b) on the right (i.e. lower tidal damping). Planets d, e, and f, are represented with dots, plus signs, and triangles, respectively. The cases studied from Figures \ref{fig:runf18_gamma_dot_zoom} and \ref{fig:runf13} are enclosed in a square and a circle, respectively. Higher variations in $\gamma$ suggest larger, or even complete, stellar coverage over time on the planet. We see clearly that there is higher probability of large variations in $\gamma$ beyond a certain threshold in the standard deviation of $\dot{n}$. We also can see that circulation becomes more common for lower tidal damping strengths (higher $Q$). Years are used for units with time.}
		\label{fig:stddevgamma_vs_forcing}
	\end{figure*}

%
%

	\subsection{Correlations}
		
	We can attempt to characterize the results of all of our 108 planetary history integrations in terms of the variance of $\gamma$ and $\dot{\gamma}$. This can be difficult given the changing nature of individual simulations, but higher standard deviation of $\dot{\gamma}$ about $0$ (i.e. the \textit{root mean squared}), and higher standard deviation of $|\gamma|$ about its mean, would generally indicate cases of higher amplitude librations, or else even complete circulation that would result in effective stellar days. We denote our measures of the variance of $\gamma$ and $\dot{\gamma}$ as $\sigma(\gamma)$ and $\sigma(\dot{\gamma})$, respectively.
	
	We can also attempt to characterize the REBOUND simulations via terms that we would expect to have the strongest effects on the evolution of $\gamma$. In our previous work, we could largely describe the driving on the spin argument $\gamma$ from the varying mean motion in terms of both the frequency and of amplitude or strength of the mean motion variations. In this work, the actual behavior of the driving is clearly more complex, but we can still attempt to largely characterize the simulations in terms of a frequency and of an amplitude of the driving. In reality, the chaotic behavior would result in varied amplitudes and multiple frequencies. But we can expect, and largely observe, a dominant frequency manifesting along with a typical amplitude.
	
	We will therefore characterize the typical amplitude of the forcing as the standard deviation of $\dot{n}$, denoted $\sigma(\dot{n})$. We also take a Fourier Transform to retrieve a primary frequency of oscillation from each of the REBOUND simulations, which we denote as $\omega_M$. As these values change depending on the REBOUND simulation, we can then relate these terms related to the driving to trends observed in the evolution of the spin argument $\gamma$. Finally, as we considered different forcing strengths and frequencies, we must also consider damping strengths. In this paper we consider two scenarios: strong damping with tidal $Q = 10$, and weak damping with $Q=100$.
	
	In Figure \ref{fig:spin_response_vs_forcing}, we demonstrate how variations in $\dot{\gamma}$ respond to different levels of forcing by plotting $\sigma(\dot{\gamma})$ versus a proxy for the amplitude of the forcing term in equation \ref{eqn:full}, $\sigma(\dot{n}) / \omega_S^2$, for different tidal damping strengths, $Q = 10$ and $Q = 100$, respectively. We normalize $\sigma(\dot{n})$ to $\omega_S^2$, as this nondimensionalizes $\sigma(\dot{n})$ to the torques on the spin and allows us to plot results from each of planets d, e, and f, on the same plot in a consistent manner, with $\omega_S \simeq 3.08$ yr$^{-1}$, $2.05$ yr$^{-1}$, and $1.38$ yr$^{-1}$ for planets d, e, and f, respectively, if eccentricity is small (valid for all these simulations). We note that these plots indeed show a correlation that as variations in $\dot{n}$ grow, (i.e. the forcing strength in equation \ref{eqn:full} increases), the spin responds more strongly with higher variations in $\dot{\gamma}$. There is also a clear dependence on tidal damping strengths, with higher tidal damping strength ($Q = 10$) suppressing variations in $\dot{\gamma}$ as compared to lower tidal damping strength ($Q = 100$). We also note that each of these plots, while manifesting a correlation of $\dot{\gamma}$ with $\dot{n}$, have a quite a bit of scatter. This is expected due to different dependencies with how the spin argument responds to the frequencies of variations in the mean motion, which adds many complexities due to the chaotic nature of the mean motion evolution in many of REBOUND integrations of the orbital parameters.
	
	As we are largely interested in how the behavior of the spin affects the total stellar coverage of these planets, we also plot variations in the spin argument $\gamma$, which we denote with $\sigma(\gamma)$, versus $\sigma(\dot{n}) / \omega_S^2$ in Figure \ref{fig:stddevgamma_vs_forcing} for different tidal factor assumptions, $Q = 10$ and $Q=100$. This demonstrates how much the substellar point on these planets is varying in longitude over the course of our integrations. We expect $\sigma(\gamma)$ to be about equal to zero for purely synchronous cases, and $\sigma(\gamma) = \pi / 2$ for cases where the spin exhibits full, regular, stellar days over the entire integration. In Figure \ref{fig:stddevgamma_vs_forcing} we notice that many simulations have planets effectively still experiencing spin-orbit synchronicity with low variance in $\gamma$ and low forcing strength. As the forcing strength increases, the variance in $\gamma$ also tends to increase, depicting higher amplitude librations in the spin. We also notice a clustering at higher forcing strengths where the $\gamma$ variations are much stronger.  This cluster represents cases wherein $\gamma$ is undergoing full circulation at least a large fraction of the time, or where the amplitudes of libration are large enough to cause switching between libration of $\gamma$ about $0$ and about $\pi$. This depicts how there is a certain threshold in libration amplitude beyond which circulation becomes much more likely, and causes this clustering at larger $\sigma(\gamma)$. 
	
	Finally, we also note that there are likely to be larger variations in $\gamma$ for lower tidal damping strengths (i.e. larger tidal factor $Q$), as we would still expect very low variance in $\gamma$ in the limit of very high tidal damping strength.

%
	
	\subsection{Best-Fit for Real System}

\begin{figure}
		\centering
		\includegraphics[width=1.0\linewidth]{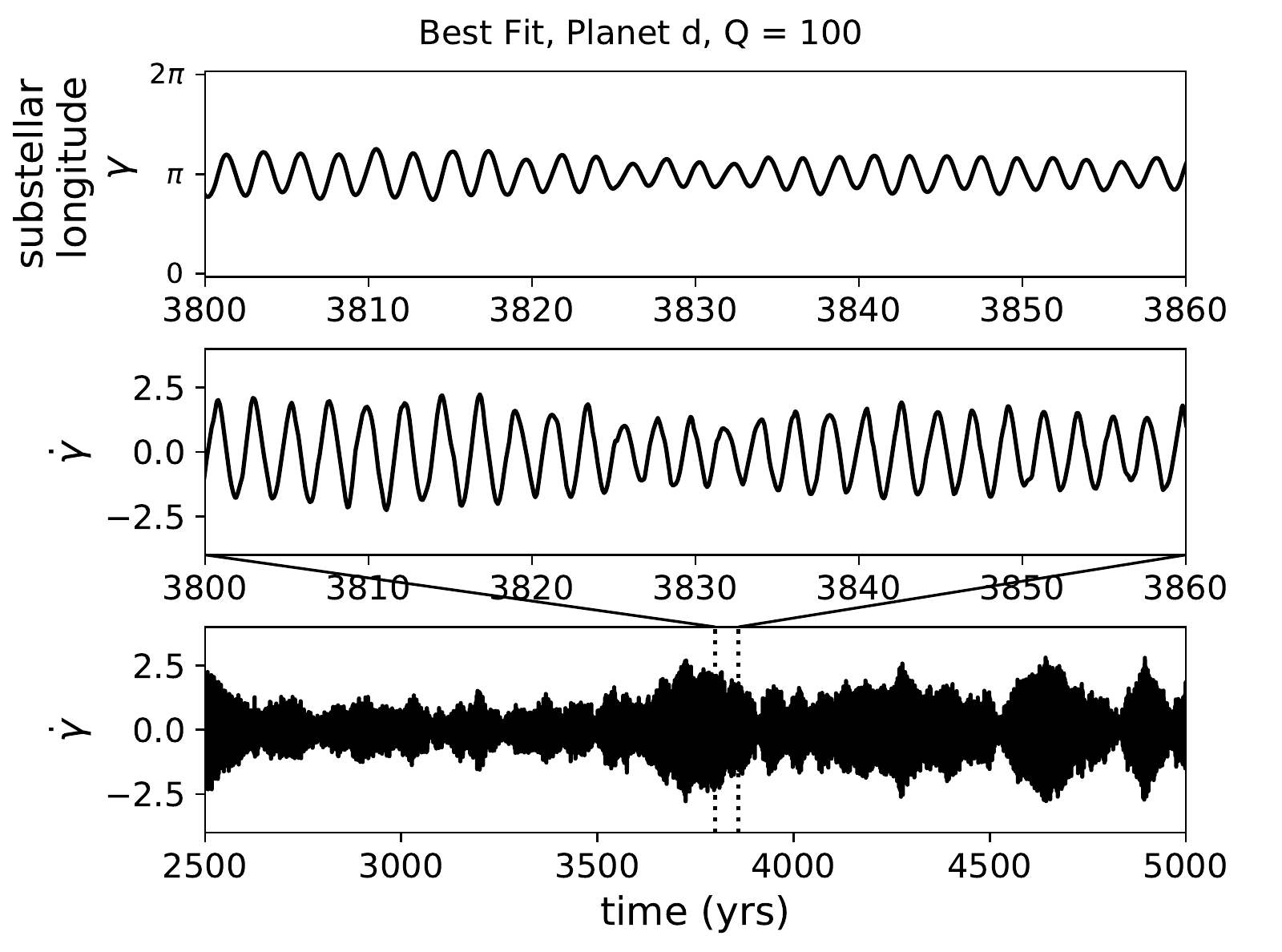}
		\caption{Best-fit results for planet d, assuming tidal factor $Q = 100$. Bottom panel shows evolution of $\dot{\gamma}$. Upper panels zoom in to show the behavior of $\gamma$ and $\dot{\gamma}$ on a shorter timescales, where we see high-amplitude libration in the spin.}
		\label{fig:best_planetd}
	\end{figure}

\begin{figure}
		\centering
		\includegraphics[width=1.0\linewidth]{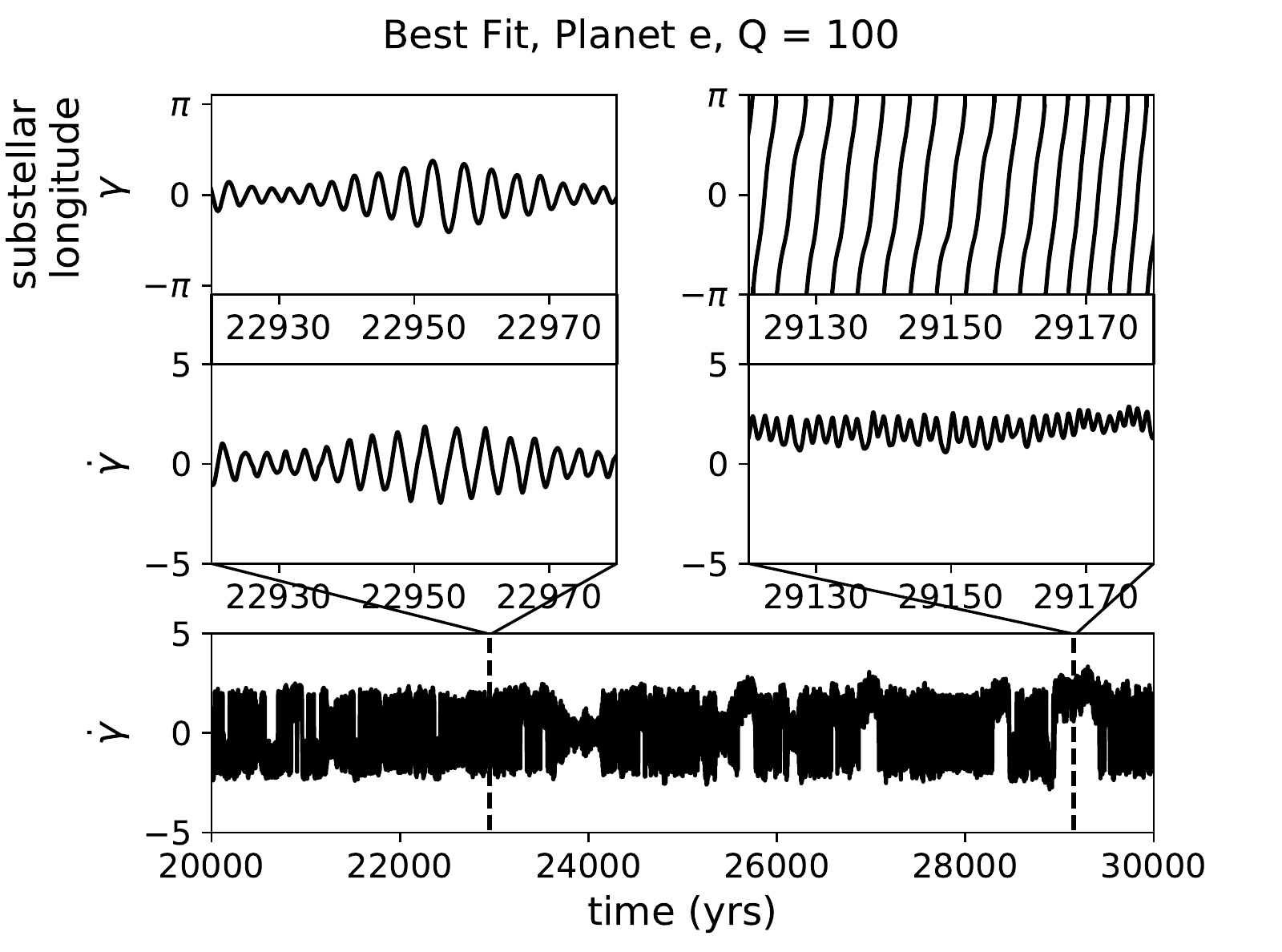}
		\caption{Best-fit results for planet e, assuming tidal factor $Q = 100$. Bottom panel shows evolution of $\dot{\gamma}$. Upper panels zoom in to show the behavior of $\gamma$ and $\dot{\gamma}$ at different times, where we see libration on the left and circulation on the right.}
		\label{fig:best_planete}
	\end{figure}

\begin{figure}
		\centering
		\includegraphics[width=1.0\linewidth]{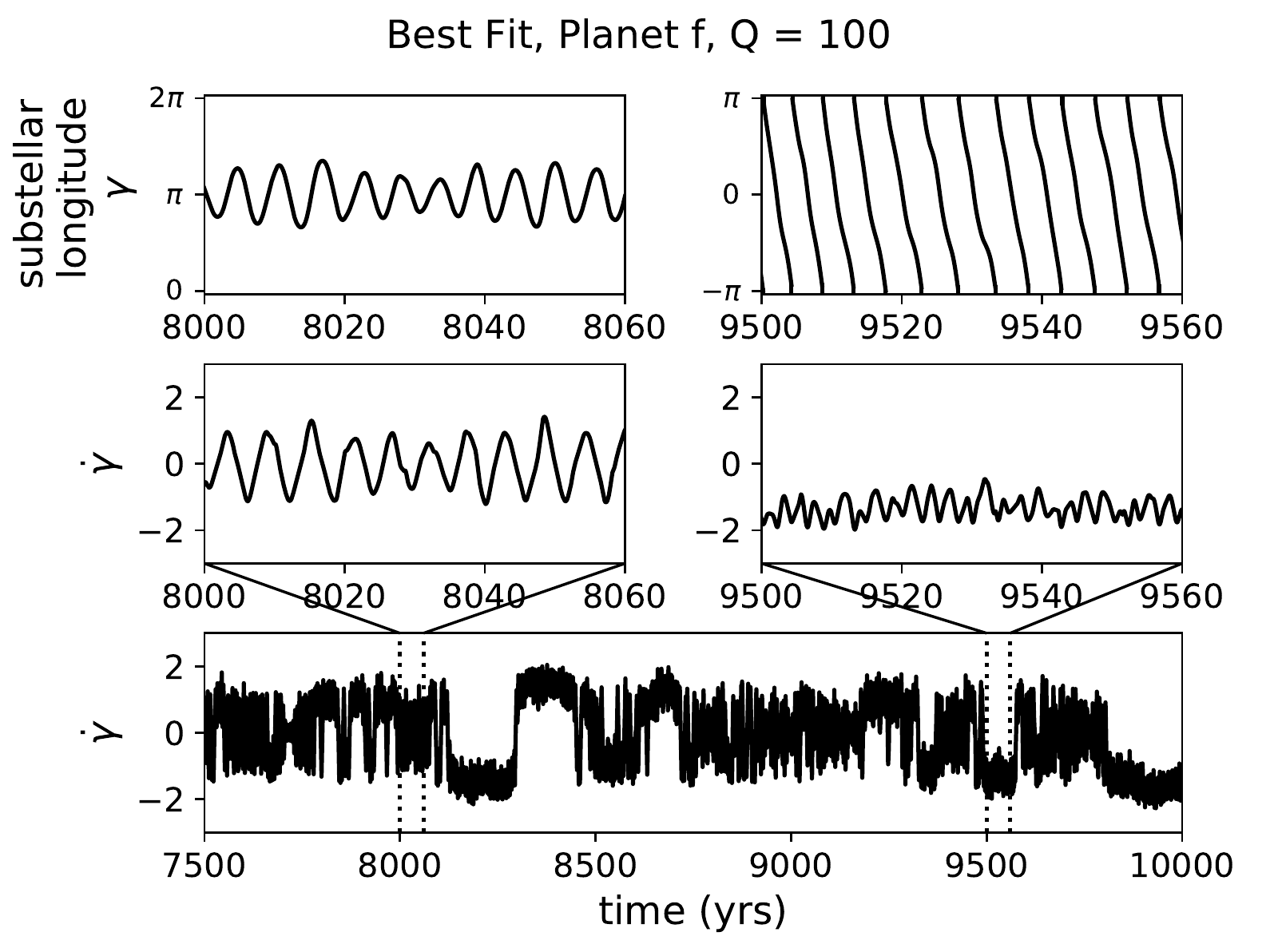}
		\caption{Best-fit results for planet f, assuming tidal factor $Q = 100$. Bottom panel shows evolution of $\dot{\gamma}$. Upper panels zoom in to show the behavior of $\gamma$ and $\dot{\gamma}$ at different times, where we see libration on the left and circulation on the right.}
		\label{fig:best_planetf}
	\end{figure}

	While our ultimate goal is to characterize how the spin of a planet may respond to a variety of possible dynamical behaviors, we may also wish to know what is likely for the real TRAPPIST-1 system. We therefore reproduced the best-fit orbital solutions from \cite{GSG18} and input the orbital histories into our spin model using the same assumptions outlined in \S \ref{setup} and with tidal factor $Q = 100$. Our results are shown for planets d, e, and f in Figures \ref{fig:best_planetd}, \ref{fig:best_planete}, and \ref{fig:best_planetf}. We can see that, indeed, the spins of each planet in the real system may respond very strongly to the mutual interactions with the other planets. We find high amplitude libration in planet d, with switching among circulation and libration in planets e and f. The spin of planet f in particular seems likeliest to respond to the strongest to the mutual interactions. We also note that, while slightly subduing the variations in $\gamma$, assuming $Q = 10$ does not drastically change our qualitative results.
	
	While these results can give a strong qualitative sense of how the spins of the planets may behave, we note that, due to its chaotic nature, exact solutions for the spin cannot predicted. However, the statistical distribution of behaviors over nearly identical initial conditions can be well defined, and we can posit that these planets may likely be effectively non-synchronous, with a high amplitude libration in the spin and with potential for circulation at certain times, resulting in more stellar coverage over their surfaces. Thus, a more detailed study could be valuable as more data becomes available and the orbits further refined.
	
	\begin{figure*}
		\centering
		\includegraphics[width=1.0\linewidth]{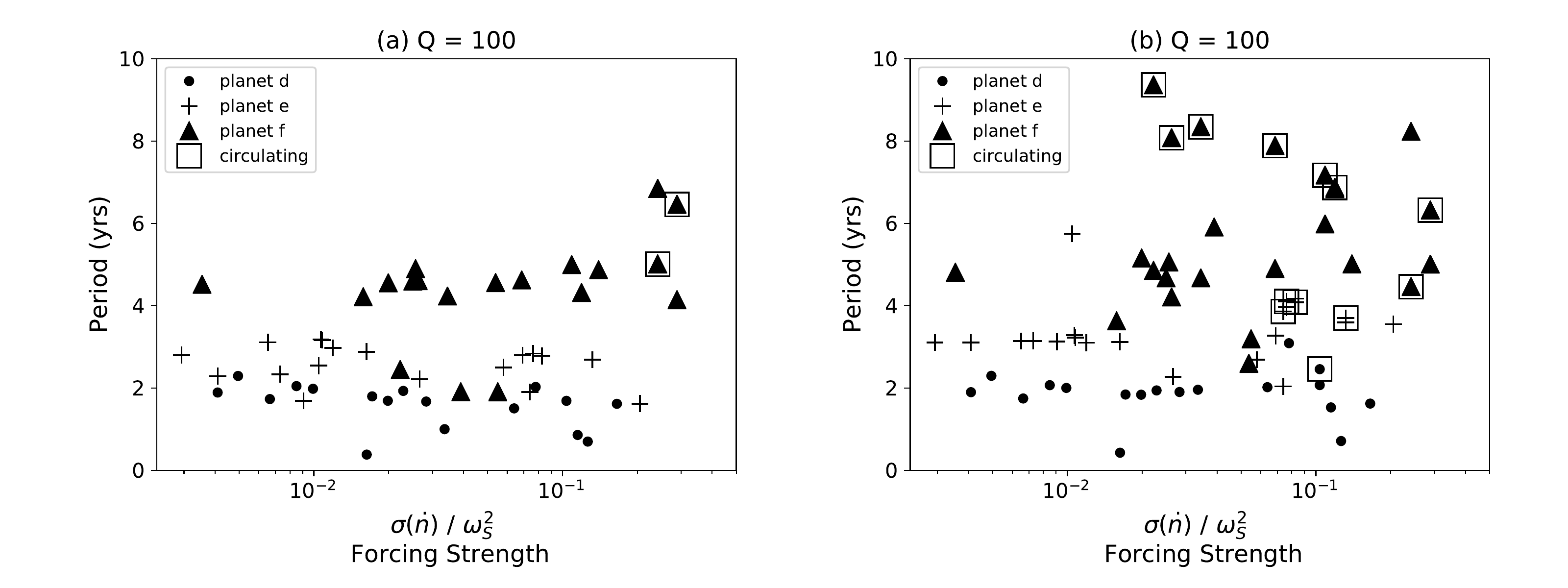}
		\caption{Periods of libration or circulation in $\gamma$ plotted against variations in $\dot{n}$ normalized to $\omega_S^2$ for our integrations with tidal factor $Q = 10$ in panel (a) on the left and tidal factor $Q = 100$ in panel (b) on the right. Integrations for planets d, e, and f, are represented by dots, plus signs, and triangles, respectively. A square around one of these markers indicates a circulating timescale for the respective planet. We find that the timescales are all on the order of a few years, with higher likelihood for periods of circulation for larger $Q$. We also note that circulation in $\gamma$ is most likely in planet f, and least likely in planet d.}
		\label{fig:periods}
	\end{figure*}

	\subsection{Timescales \label{timescales}}
	
	There are two timescales which are of primary concern in our results for the spin evolution: one is the libration or circulation timescales, and the other is the duration of quasi-stable states, that we observe in many of our simulations. These timescales both would have profound implications for climate and thus for habitability. 
	
	The duration of quasi-stable states varies among our simulations from just a few thousand years to hundreds of thousands of years. In the example presented in Figure \ref{fig:runf18_gamma_dot_zoom}, we determine that the circulating and librating states endure on timescales on the order of $10^4$ years. The planet would undergo long periods of moderate-amplitude librations of its spin argument $\gamma$, while occasionally switching to other quasi-stable states, some where spin argument $\gamma$ can ``flip'' such that the side that was previously the ``night side'' becomes the ``day side'' and vice versa, and other states where the spin argument $\gamma$ can fully circulate for a few thousand years.
	
	The other important timescale to note here is the period of the libration or circulation while in a quasi-stable state. Figure \ref{fig:periods} depicts the average timescale for libration and circulation for each of our integrations for tidal factors $Q = 10$ and $Q = 100$. Here we see typical timescales on the order of a few years for both librating and circulating cases. This implies, for the circulating cases, full stellar days, whererin the planet makes a full rotation in its co-rotating frame, which last on the order of a few years (while the system lasts in such a circulating quasi-stable state). We note that these stellar days last far longer than the orbital period of the planet, with the planet's spin still being effectively synchronous over one orbit. We also note that circulating behavior is more common for lower tidal damping strengths (corresponding to larger tidal factor $Q$), and also appears to be more common for the outer-most planet we simulated, planet f, and less common for inner-most simulated planet, planet d. This would also be expected, as the tidal damping strength, as well as depending on $Q$, also depends strongly on distance from the central star, with tidal damping increasing with decreased distance from the star. We note that circulation is still a possibility even for planet d under the right circumstances.

	\section{Discussion} \label{discussion}
	
	Our model depicts how much more complicated the spin states of planets in compact, multi-planet systems can be than what one might naively infer from simple tidal theory. The chaotic configuration of a system such as TRAPPIST-1 can lead to a chaotic driving of the spin argument $\gamma$, resulting in a variety of possible chaotic behaviors in the spin. Some simulations exhibit small or moderate amplitude librations of $\gamma$, while other depict full circulation, with many others exhibiting switches among different quasi-stable states. This can clearly have dramatic potential consequences on climate, and thus on habitability prospects. 
	
	From \S \ref{timescales}, we noted that the spin timescales for libration or circulation are on the order of a few years, as also seen in Figure \ref{fig:periods} . These timescales are shorter than the expected atmospheric response time of $\lesssim 10$ years for Earth-like planets as calculated by \cite{SMS}, and thus can help prevent atmospheric collapse on the night-side of the planet if the spin is circulating or undergoing high-amplitude libration.
	
	We also note that state switching could also be likely, wherein the spin state changes behavior by abruptly transitioning from librating to circulating, circulating to librating, or else the planet swapping day and night sides for librating cases (e.g. switching from librating about $\gamma$ at $0$ to librating about $\pi$ on the opposite face of the planet). The longevity of these quasi-stable states can be on the order of a few thousand years, though the time it takes to switch from one state to another is on the order of tens of years. This can result in relatively stable climate states that last for a few thousand years, but that can then intermittently switch.
	
	That the libration and circulation timescales are less than the expected atmospheric response times of these planets would seem to bode well for habitability prospects on these planets, if the response of the spin to the driving of mean motion variations caused by neighboring planets is strong enough to produce high-amplitude librations in $\gamma$, or even complete circulation. The effects of the intermittent changes in climate, due to switches among quasi-stable spin states, is less certain. The time for the switches to occur would provide a period of transition on the order of tens of years from one quasi-stable climate period to another, likely giving the atmosphere enough time to respond gradually enough. More work is necessary to fully explore the effects on climate and the resultant effects on potential habitability.

	With the new Transiting Exoplanet Survey Satellite (\textit{TESS}), which aims to perform a near-full sky survey designed for detecting small exoplanets in tight configurations, there comes the opportunity of discovering many similar systems in which our model may be applicable. We should note, however, that even given an optimistic scenario of getting many relevant analog systems from \textit{TESS}, we wouldn't expect to be able to observe any spin transitions occuring. Rather, we would observe a series of snapshots of the system.
	
	\section{Conclusion} \label{conclusion}
	
	We have applied the spin model introduced in \cite{VH17} more completely to different, stable configurations of the TRAPPIST-1 system. Studying different possible configurations and orbital histories of the system allowed us to more fully demonstrate possible spin behaviors for the system or systems like it. Our model exhibits a range of possible responses of the spin behavior of a planet to the influences of nearby companions in resonant or near-resonant configurations.
	
	We find that, for a system such as TRAPPIST-1, which contains seven planets in a long resonant chain, the resultant behavior of the spin can be varied and unpredictable. Such a multiple and compact system can have chaotic evolutions of orbital parameters, which could then cause chaotic evolutions of the spin. There can exist, however, quasi-stable spin states which remain stable for thousands of years before abruptly switching into a different quasi-stable state. Other simulations depict spin evolutions which are far less chaotic and are long-term stable (see Figure \ref{fig:runf13}, which depicts a spin-state which is long-term stable and exhibits moderate-amplitude librations in the spin argument $\gamma$).
	
	These varied results suggest varied consequences for climate and habitability on these planets. It is often assumed that these planets orbiting so close to their host star must be tidally locked into synchronously rotating spin state, but our work here demonstrates that this isn't necessarily the case. We show that, depending on factors such as the strength of variations in mean motion due to the presence of nearby planetary companions, the spin of such planets may exhibit libration or even complete circulation, with timescales of libration and circulation on the order of years and shorter than the expected atmospheric response time. These results therefore should prompt further and more detailed investigation climate effects.
	
	\section*{Acknowledgments}
	
	The authors thank the anonymous referee for their comments, which helped greatly improve upon our original manuscript. The authors also thank Simon Grimm and Brice-Olivier Demory for providing us with the best-fit orbital histories for the TRAPPIST-1 planets from \cite{GSG18}.

	\bibliographystyle{mnras}
	\bibliography{trapp_spins}

	\appendix
	
	\section{REBOUND Samples}\label{appendix}
	
	We sampled from different initial conditions of REBOUND simulations of the orbital parameters of TRAPPIST-1. We chose to sample among a range of different orbital behaviors. Namely, we sampled among simulations with behaviors spanning a range in the forcing strength depicted in equation \ref{eqn:full} (i.e. simulations spanning a range of different variances in mean motion among the planets), from negligible to strong (see horizontal axes of Figures \ref{fig:spin_response_vs_forcing}, \ref{fig:stddevgamma_vs_forcing}, and \ref{fig:periods}). REBOUND simulations sampled by file name are listed below and can be found at \url{https://zenodo.org/record/496153}:\\ \\ IC111K1.6763e+02mag3.2152e-03.bin IC100K1.2213e+02mag6.8408e-03.bin IC125K1.0315e+02mag1.4880e-03.bin IC235K7.2640e+02mag3.6993e-02.bin IC123K2.4714e+02mag7.2180e-03.bin IC119K4.9624e+02mag3.1153e-02.bin IC73K1.9292e+02mag4.1292e-02.bin IC190K3.0921e+02mag7.6131e-02.bin IC163K2.3874e+02mag2.6440e-02.bin IC98K2.9143e+02mag5.0300e-02.bin IC290K3.5223e+01mag6.1289e-03.bin IC294K2.1680e+02mag2.5643e-03.bin IC93K1.6320e+02mag9.3872e-02.bin IC101K1.0784e+02mag5.1523e-02.bin IC162K4.3853e+01mag1.8504e-01.bin IC239K4.1655e+01mag2.4514e-03.bin IC240K3.8638e+01mag1.9549e-02.bin IC70K7.1608e+02mag4.1427e-01.bin

\end{document}